\documentclass[aps,prl,twocolumn,superscriptaddress]{revtex4-2}
\usepackage{amssymb}
\usepackage{graphicx}
\usepackage{amsmath}
\usepackage{amsthm}
\usepackage{amsfonts}
\usepackage{array}
\usepackage{multirow}
\usepackage{color}
\usepackage{esint}
\usepackage{bm}
\usepackage{color}
\usepackage{bbm}
\usepackage{hyperref}
\usepackage{titlesec}
\usepackage{changes}
\usepackage{subfigure}
\usepackage{physics}
\usepackage{siunitx}
\usepackage[english]{babel}
\usepackage[utf8]{inputenc} 

\newcommand{\lettersection}[1]{\emph{\color{cyan} #1.---}}

\begin{document}

\title{Coexistence of distinct mobility edges in a 1D quasiperiodic mosaic model}

\author{Xu Xia}
\affiliation{Academy of Mathematics and Systems Science, Chinese Academy of Sciences, Beijing 100190, China }
\author{Weihao Huang}
\affiliation{Department of Physics, City University of Hong Kong, Kowloon, Hong Kong SAR}
\author{Ke Huang}
\affiliation{Department of Physics, City University of Hong Kong, Kowloon, Hong Kong SAR}
\author{Xiaolong Deng}
\affiliation{Leibniz Supercomputing Centre, Boltzmannstraße 1, 85748 Garching bei München, Germany}
\author{Xiao Li}
\email{xiao.li@cityu.edu.hk}
\affiliation{Department of Physics, City University of Hong Kong, Kowloon, Hong Kong SAR}
\date{\today}

\begin{abstract}
We introduce a one-dimensional quasiperiodic mosaic model with analytically solvable mobility edges that exhibit different phase transitions depending on the system parameters. 
Specifically, by combining mosaic quasiperiodic next-nearest-neighbor hoppings and quasiperiodic on-site potentials, we rigorously demonstrate the existence of two distinct types of mobility edges: those separating extended and critical states, and those separating extended and localized states. 
Using Avila's global theory, we derive exact analytical expressions for these mobility edges and determine the parameter regimes where each type dominates. 
Our numerical calculations confirm these analytical results through fractal dimension analysis. Furthermore, we propose an experimentally feasible scheme to realize this model using Bose-Einstein condensates in optical lattices with engineered momentum-state transitions. 
We also investigate the effects of many-body interactions under mean-field approximation. 
Our work provides a fertile ground for studying the coexistence of different types of mobility edges in quasiperiodic systems and suggests a feasible experimental platform to observe and control these transitions. 
\end{abstract}

\maketitle

\lettersection{Introduction}
Anderson localization, the exponential localization of quantum states due to disorder~\cite{Anderson1958}, is one of the most extensively studied phenomena in quantum physics.
The phase transition between localized and extended phases in disordered systems is highly dependent on the system's dimensionality. 
In systems with three or more dimensions, a finite disorder strength is required to induce the localization transition~\cite{Anderson1979}. 
Moreover, for intermediate disorder strengths, this transition may be energy-dependent, with critical energies known as single-particle mobility edges (MEs) that distinctly separate localized from extended states~\cite{Evers2008}. 
In contrast, in one-dimensional (1D) and two-dimensional (2D) systems with random disorder, all states are localized even at infinitesimally small disorder strengths, preventing a localization transition and the formation of MEs.

Besides the random disorder, quasiperiodic potential can also induce localization transitions in 1D systems. 
Such models have attracted significant attention not only due to their experimental realization in ultracold atomic gases~\cite{Roati2008, Fallani2007, Bloch1, Bordia, Bloch2, Modugno2014, Bloch3, Bloch4} but also because they exhibit interesting phenomena such as the presence of extended-localized transitions and MEs even in 1D systems~\cite{Xie1988, Hashimoto1992n, Biddle2009, Biddle, Gao2019, Li2017, Bloch4, Yao2019, Ganeshan2015, Xu2020,Li2020}. 
The Aubry-Andr\'{e}-Harper (AAH) model~\cite{AA} is a prototypical example of a 1D quasiperiodic system that exhibits a localization transition. 
The AAH model has a self-duality property that leads to all states being either extended or localized, with no SPMEs. 
However, by breaking the self-duality of the AAH model, such as by introducing long-range hopping terms~\cite{Biddle} or additional quasiperiodic optical lattices~\cite{Li2017, Bloch4, Yao2019, Ganeshan2015, Xu2020, Li2020}, one can induce MEs and study the localization transition in 1D quasiperiodic systems. 
Despite significant theoretical progress, experimentally distinguishing critical states from extended states remains challenging, as both appear delocalized in conventional measurements. 
A fundamental question emerges: beyond the well-studied MEs separating extended and localized states, can other types of MEs exist, such as those separating critical from extended states? 
Furthermore, can a single model host multiple types of MEs simultaneously? 
These questions become even more intriguing when considering many-body interactions, which could fundamentally alter the nature and behavior of these different types of MEs.

Motivated by these questions, we introduce in this Letter an analytically solvable 1D quasiperiodic mosaic model with mosaic quasiperiodic next-nearest-neighbor (NNN) hoppings and mosaic quasiperiodic onsite potential. 
Through Avila's global theory~\cite{A4}, we analytically prove that the MEs of this model change from the ones that separate extended and critical states to the ones that separate extended and localized states as the potential strength increases. 
We numerically confirm these results by calculating the fractal dimension of the eigenstates. 
Moreover, we propose a concrete experimental scheme based on the Bose-Einstein condensate (BEC)~\cite{Roati2008} to realize this model. 
Finally, because interactions naturally exist in BEC experiments~\cite{Wang_2022, Gadway_2021}, a pertinent and interesting question is how our single-particle theory is modified by many-body interactions in the BEC experiments. 
We specifically study the effects of many-body interactions on the critical-localized transition in the ground state under the mean-field approximation.
Using finite-size scaling analysis~\cite{Yucheng_2022}, we demonstrate that many-body interactions significantly modify the nature of single-particle states: repulsive interactions ($U<0$) transform critical states into extended states and drive localized states toward criticality, while attractive interactions ($U>0$) convert critical states into localized states with localized states remaining localized. 
Additionally, we identify two types of localized states with unbroken and spontaneously broken inversion symmetry, which arise due to the competition between the quasiperiodic on-site potential and the NNN hopping. 
Our study provides a fertile ground for exploring the coexistence of different types of MEs in quasiperiodic systems and also paves the way for the experimental realization of these transitions in ultracold atomic gases.

\begin{figure}[t]
\centering
\includegraphics[width=\columnwidth]{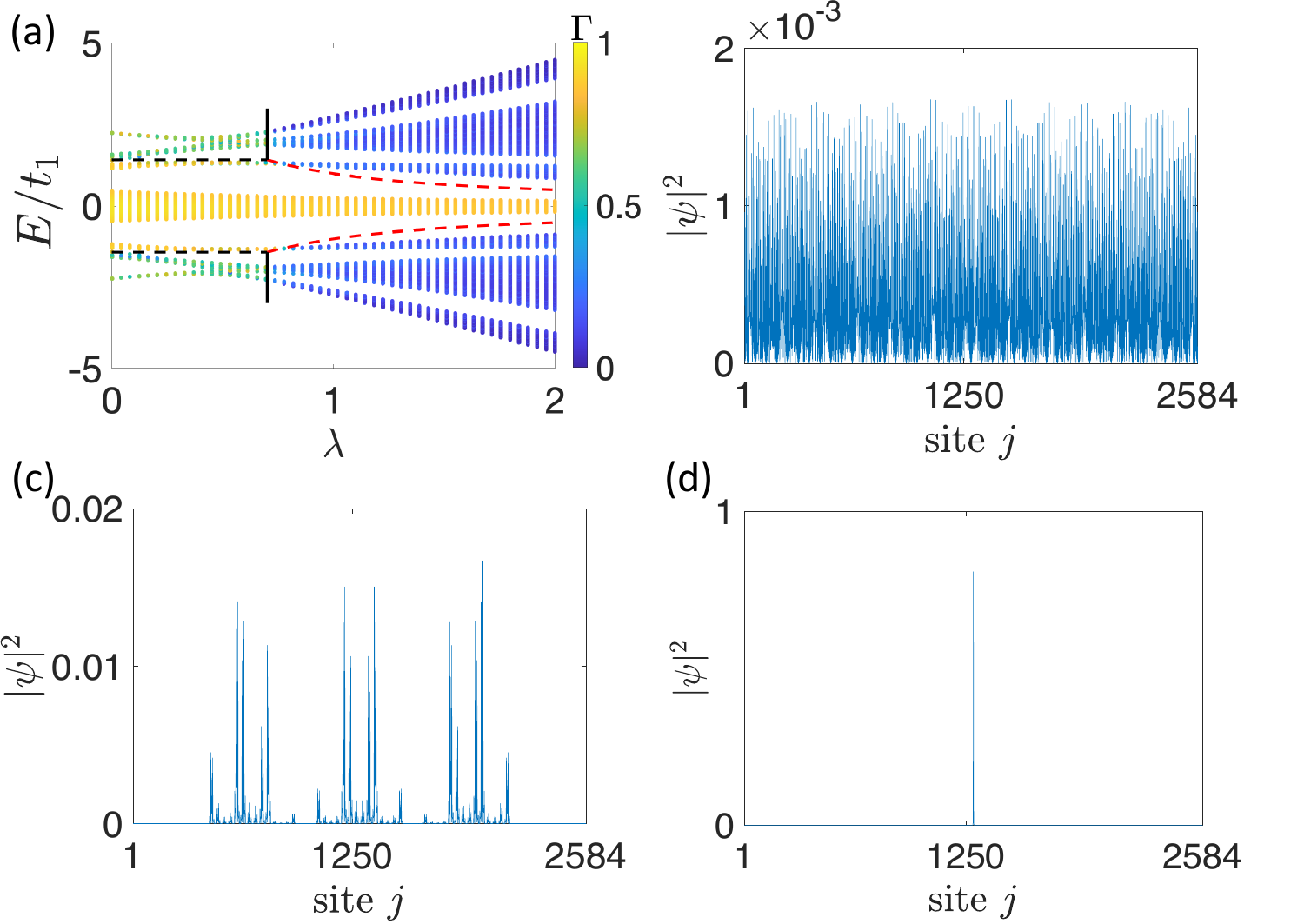}
\caption{\label{fig1}(a) Fractal dimension $\Gamma$ of different eigenstates as a function of the energy eigenvalues and quasiperiodic potential strength $\lambda$ with size $L=2584$ and $\mu=0.35$. 
The black and red dashed lines represent the AMEs and MEs, respectively. 
Here, the hopping strength $t_1$ has been set as the unit of energy. 
We also show the density distributions of the following eigenstates: 
(b) $\lambda=0.6$, $E/t_1=-0.066$, (c) $\lambda=0.6$, $E/t_1=2.1906$ and (d) $\lambda=1$, $E/t_1=2.7541$.}
\end{figure}

\lettersection{The model}
We start by describing our 1D quasiperiodic mosaic model, which is a generalization of the AAH model with quasiperiodic NNN hoppings and quasiperiodic on-site potentials, given by 
\begin{equation}
H =\sum_{j}[t_1c^{\dagger}_{j}c_{j+1}+t_2(j)c^{\dagger}_{j}c_{j+2}+\text{h.c.}]+\sum_jV_j n_j,
\label{Eq:model}
\end{equation}
where $c_{j}(c_j^\dagger)$ is the annihilation (creation) operator at site $j$, $n_{j}=c^{\dagger}_{j}c_{j}$ is the local density operator, and $t_1$ is the nearest neighbor (NN) hopping coefficient. 
Moreover, $t_2(j)$ and $V_j$ denote the quasiperiodic NNN hopping and potential amplitude, respectively, where $t_2(j)=V_j=0$ for odd $j$ and
\begin{equation}
\begin{aligned}
t_2(j)=&\ 2 \mu \cos [2 \pi\alpha(j+1)+\theta],\\
V_j=&\ 2\lambda\cos(2\pi\alpha j+\theta),
\end{aligned}
\end{equation}
for even $j$. 
Here $\theta$ is the phase offset, and $\alpha$ is an irrational number.

A central result of this Letter is that we can rigorously prove the existence of different types of MEs in this model and derive their closed-form expressions by computing the exact Lyapunov exponent (LE). 
For convenience, we use ME to denote the mobility edge that separates extended and localized states and anomalous mobility edge (AME) for the ME that separates critical and extended states. 
With this convention, we make the following statement: 
When $\lambda>2 \mu$, there are only MEs given by $E_\text{ME}$; when $0<\lambda<2 \mu$, there are only AMEs given by $E_\text{AME}$, where 
\begin{equation}
  E_\text{ME} =\pm {t_1^2}/{\lambda}, \quad 
  E_\text{AME} =\pm {t_1^2}/{(2\mu)}.
  \label{Eq:ME}
\end{equation}

Before presenting the proof of the above result, we numerically compute the ME and AME in the model by calculating the fractal dimension of the wave functions and the spatial distribution of the eigenstates. 
For this simulation, we set $t_1=1$, $\theta=0$ and $\alpha={(\sqrt{5}-1)}/{2}$. 
This $\alpha$ can be approached by the ratio of two consecutive Fibonacci numbers $F_{m}$~\cite{Kohmoto1983,Wang2016n}: 
$\alpha=\lim_{m \rightarrow \infty}\qty(F_{m-1}/F_{m})$. 
Hence, we take the system size as $L=F_{m}$ (for a large $m$), adopt the rational approximation $\alpha=F_{m-1}/F_{m}$, and impose the periodic boundary condition.  
The fractal dimension of the $j$th eigenstate is defined as 
$\Gamma_j=-\lim_{L\rightarrow\infty}\qty[\ln(\mathcal{I}^{(2)}_j)/\ln L]$, 
where 
\begin{equation}
  \mathcal{I}^{(q)}_j=\sum_{i=1}^{L}|\psi_{j}(i)|^{2q}  \quad 
  (q = 2, 3, 4, \ldots)
  \label{Eq:IPR}
\end{equation}
is the inverse participation ratio (IPR) of the $j$th eigenstate with $\psi_{j}(i)$ denoting the wave function of the $j$th eigenstate at site $i$. 
It is known that $\Gamma\to 1$ for extended states, $0<\Gamma< 1$ for critical states, and $\Gamma\to 0$ for localized states. 
We plot the eigenvalue $E$ and the fractal dimension $\Gamma$ of each eigenstate as a function of potential strength $\lambda$ in Fig.~\ref{fig1}(a). 
As expected from our analytical results, Fig.~\ref{fig1}(a) shows that the  transitions between different types of eigenstates occur precisely at the analytically derived ME and AME lines in Eq.~\eqref{Eq:ME}.  
This classification is further confirmed by examining the spatial density distributions of representative eigenstates in Fig.~\ref{fig1}. 
For $\lambda>2\mu$, we observe two distinct behaviors separated by the MEs at $\abs{E}={t_1^2}/{\lambda}$ (red dashed lines): states with $\abs{E}>{t_1^2}/{\lambda}$ are exponentially localized on a single site [Fig.~\ref{fig1}(d)], while states with $\abs{E}<{t_1^2}/{\lambda}$ spread over the entire system [Fig.~\ref{fig1}(b)]. 
For $0<\lambda<2\mu$, the AMEs at $\abs{E}={t_1^2}/{2\mu}$ (black dashed lines) separate extended states ($\abs{E}<{t_1^2}/{2\mu}$) from critical states ($\abs{E}>{t_1^2}/{2\mu}$), with an example of the latter shown in Fig.~\ref{fig1}(c).

The physical origin of these distinct mobility edges can be understood from the model's structure: the incommensurate NNN hopping between even sites induces critical states~\cite{liuxiongjun_2023}, while the vanishing onsite energies at odd sites combined with the NN hopping maintain extended states in the middle of the spectrum even for large potential strengths~\cite{liuxiongjun_2020}. 
The competition between these mechanisms results in the different types of MEs.

\lettersection{Mathematical derivation of the ME} 
Here, we provide the analytical derivation for the MEs by computing the LE. 
The eigenstates of Eq.~\eqref{Eq:model} can be equivalently expressed in terms of the transfer matrix:
\begin{equation}
T_m(\theta)\left(\psi_j,\psi_{j-2}\right)^\intercal=\left(\psi_{j+2},\psi_{j}\right)^\intercal.
\label{Eq:chara_eq}
\end{equation}
Since wave functions on odd sites are fully determined by those on even sites, we focus only on the latter.
Denote the transfer matrix at site $j=2m$ by $T_m(\theta)$, and the product of transfer matrices as 
$\mathcal{T}_m=T_mT_{m-1}\cdots T_0$. 
From Eq.~\eqref{Eq:model}, we obtain the transfer matrix on even sites as 
$T_m(\theta)=\qty[t_2(j-2)+t_1^2 / E]^{-1}B_m(\theta)$, where
\begin{align*}
B_m(\theta)=
\mqty[
  E-2t_1^2/E-V_{j} & -t_2(j)-t_1^2/E\\
t_2(j-2)+t_1^2/E & 0
].
\end{align*}
The LE can be computed as 
\begin{align}
\gamma_{\epsilon}(E)=\lim_{m\rightarrow \infty}\frac{1}{m} \int \ln  \|\mathcal{T}_m(\theta +i \epsilon)\| \dd{\theta}.
\end{align}
By applying Avila's global theory of one-frequency analytical $SL(2,\mathbb{R})$ cocycle~\cite{A4}, we obtain the LE of an eigenstate as 
$\gamma_0(E)= \max\{\alpha(E), 0\}$, where 
\begin{align}
  \alpha(E)=\ln \frac{\abs{E}\left(\lambda+\sqrt{\lambda^2-4 \mu^2}\right)}{t_1^2 + \sqrt{t_1^4-4 E^2 \mu^2}}. 
\end{align}
When $\lambda>2 \mu$, the separation line is $\alpha(E)=0.$ 
Hence, the MEs are $\abs{E}=t_1^2/{\lambda}$. 
Therefore, when $\abs{E}>t_1^2/{\lambda}$, we have $\gamma_0(E)>0$ indicating localized states. 
In contrast, $\abs{E}<t_1^2/{\lambda}$, we have $\gamma_0(E)=0$ indicating extended states. 
When $0<\lambda<2 \mu$ and $\abs{E}> t_1^2/{2 \mu}$, however, the transfer matrix exhibits singular properties because the element $t_2(j-2)+t_1^2 / E$ can vanish at certain sites. 
This singularity prevents the existence of an absolutely continuous spectrum, resulting in only singular continuous spectrum, which characterizes critical states. 
Thus, when $0<\lambda<2 \mu$, the AMEs are given by $
E=\pm t_1^2/{2 \mu}$. 
Further details of the mathematical derivation are shown in the Supplemental Materials (SM)~\cite{SM}.

\begin{figure}[t]
\centering
\includegraphics[width=\columnwidth]{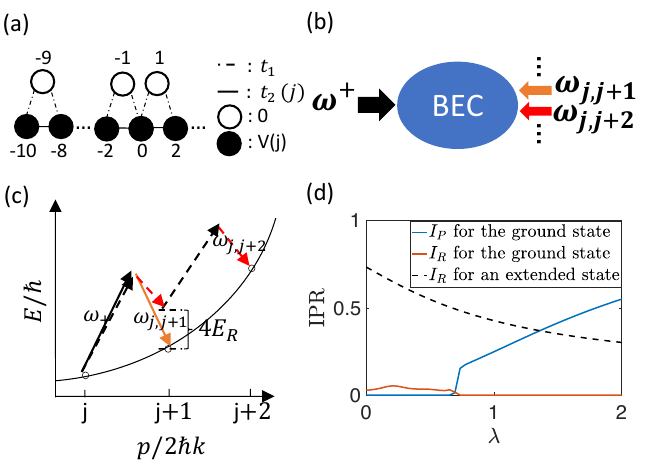}
\caption{\label{fig2}
(a) A schematic setup of the experimental realization of the 1D quasiperiodic mosaic model in Eq.~\eqref{Eq:model} using ultracold Cesium atoms. 
Here we assume that the system contains $21$ different momentum states (as lattice sites).   
(b) Two counterpropagating laser beams are applied to the BEC: one with a single frequency and one with multiple frequencies, which together drive controlled momentum transitions between atomic states.
(c) The first-order (solid arrows) and second-order (dashed arrows) Bragg transitions drive the NN hoppings and NNN hoppings, respectively, between the momentum states. Here $E_R=\hbar^2k^2/2M_{\text{Cs}}$ is the recoil energy of the Cesium atoms. 
(d) Numerical simulations of the IPR for the ground state as a function of quasiperiodic potential strength $\lambda$ in both momentum space ($I_P$) and real space ($I_R$), validating our experimental detection protocol. For comparison, the IPR of a reference extended state is also shown. To minimize finite-size effects, we use $q=6$ in the calculation of both $I_P$ and $I_R$ according to Eq.~\eqref{Eq:IPR}. The real-space wavefunctions are obtained through Fourier transformation of the momentum-space wavefunctions. 
The simulation parameters are $L=21$ sites, $\mu=0.35t_1$, and phase offset $\theta=\pi$.
}
\end{figure}

\lettersection{Experimental realization}
Having established the existence of different types of MEs in our model, we now propose an experimental scheme to realize this model using ultracold atoms in optical lattices. 
The schematic setup is illustrated in Fig.~\ref{fig2}(a) for a system with $21$ sites. 
The main difficulty in the experimental realization of this model lies in the engineering of the quasiperiodic NNN hopping. 
In this regard, the momentum-space lattice built from ultracold atoms~\cite{Wang_2022} provides a feasible solution, as shown in Fig.~\ref{fig2}(b). 
The BEC cloud is subject to two counterpropagating laser beams: one with a single frequency of $\omega^+=2\pi c/\lambda$ where $c$ is the speed of light and $\lambda=\SI{1064}{nm}$, and the other with multiple frequencies. 
Initially, the atoms are prepared into the momentum $\hbar k=0$ state. 
Turning on these lasers would drive a series of Bragg transitions that only allows quantized atomic momentum changes in units of $2\hbar k$. 
The NN hoppings between the sites (related to the atomic momentum $p_j=2j\hbar k$) are realized by the first-order, two-photon Bragg transition, i.e., absorbing one photon from the beam with frequency $\omega_+$ and emitting one photon into the multifrequency beam.
Similarly, the NNN hoppings are realized by the second-order, four-photon Bragg transition~\cite{Gadway2018}, where two photons are absorbed from the single-frequency beam, and two photons are emitted into the multifrequency beam, as shown in Fig.~\ref{fig2}(c). 
Due to the quadratic dispersion, the energy difference between any pair of momentum states is unique. 
Thus, one can individually manipulate the strength and phase of the corresponding frequency beam to control the NN and NNN hoppings. 
Finally, the quasiperiodic potential can be engineered by setting the difference between the frequencies of the single-frequency beam and the multi-frequency beams as $2\lambda\cos{(2\omega\pi j+\theta)}$. 
Further details of the experimental setup is shown in the SM~\cite{SM}. 

\begin{table}[!]
\centering
\begin{tabular}{c|c|c}
\hline\hline
 & Momentum-space IPR ($I_P$) & Real-space IPR ($I_R$) \\
\hline
Extended & $I_P\sim 0$ & Finite \\
Critical & $I_P\sim 0$ & $I_R\sim0$ \\
Localized & Finite & $I_R\sim0$ \\
\hline\hline
\end{tabular}
\caption{Comparison of inverse participation ratio (IPR) characteristics in momentum space and real space for extended, critical, and localized states.}
\label{tab:IPR}
\end{table}

A hallmark feature of our model [Eq.~\eqref{Eq:model}] is that its ground state undergoes a critical-to-localized transition precisely at $\lambda = 2\mu$, providing an ideal platform for experimental detection of critical states.
We propose an experimental protocol to observe this transition, beginning with the initialization of the BEC cloud at site $j=0$ with phase $\theta=\pi$ in the quasiperiodic potential. This configuration serves as the ground state of the system without hopping terms.
To prepare the complete ground state, we adiabatically introduce the hopping terms by linearly ramping the NN hopping strength from zero to $t_1$ and the NNN hopping strength from zero to $t_2(j)=2\mu t_1\cos{[2\pi\omega (j+1)+\pi]}$ on even sites, followed by a holding period.
Time-of-flight measurements can then determine the atomic momentum distribution, allowing calculation of the momentum-space IPR ($I_P$), which effectively distinguishes localized states ($I_P \sim \text{finite}$) from critical states ($I_P\sim 0$). 
To further differentiate between critical and extended states, we propose measuring the real-space distribution to calculate the real-space IPR ($I_R$), since critical states in the momentum space remain critical in real space while extended states in momentum space become localized in real space, resulting in distinct $I_R$ values as summarized in Table~\ref{tab:IPR}.
Figure~\ref{fig2}(d) presents numerical simulations in an $L=21$ system validating this experimental approach.

\lettersection{Effects of many-body interactions}
In realistic BEC experiments, interactions between atoms can be easily turned on by introducing $s$-wave collisions between atoms~\cite{Wang_2022, Gadway_2021}. 
To study how these interactions affect the critical-localized transition in our quasiperiodic mosaic model, we employ the mean-field Gross-Pitaevskii equation (GPE), where interactions are treated as an effective density-dependent potential. 
Specifically, the GPE reads as 
$i\hbar\dot\psi=H_{\text{MF}}\psi$, where 
\begin{align}
  H_{\text{MF}} = H-U\sum_j|\psi_j|^2n_j. 
\end{align}
In the above equation, $\abs{\psi_j}^2$ is the density probability of the wavefunction on each site, and $U$ is the effective interaction strength. 
We choose $\theta=\pi$ and the same setup as the previous experimental proposal. 
The open boundary condition is considered. 

\begin{figure}[t]
  \includegraphics[width=\columnwidth]{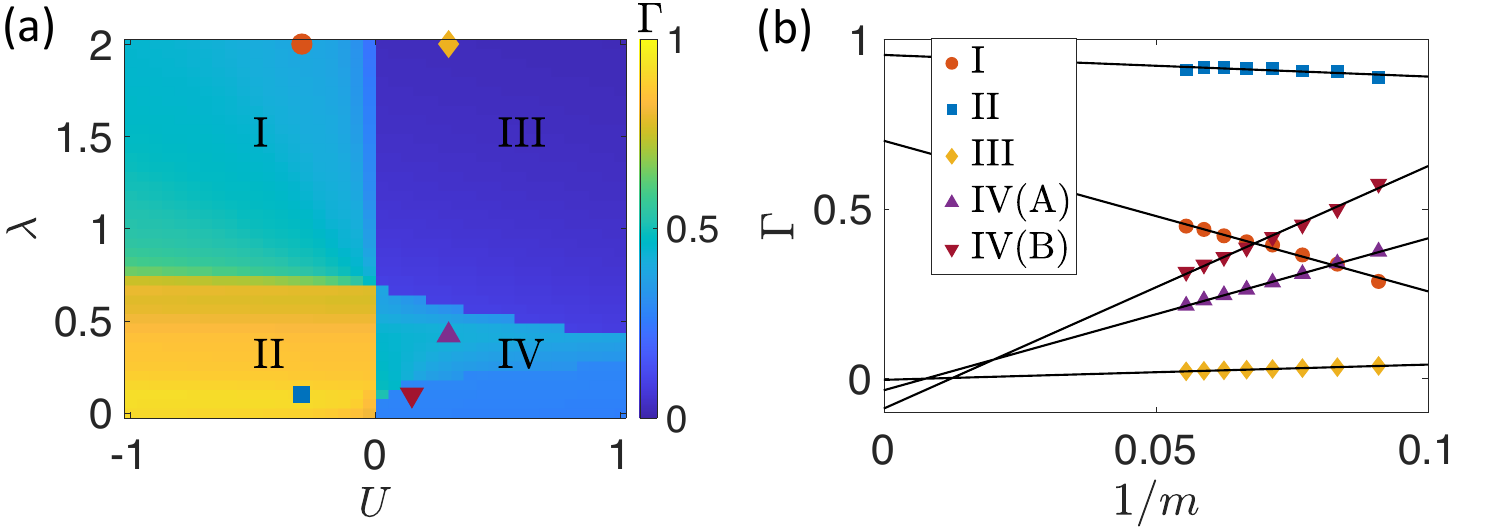}
\caption{
\label{fig3}
(a) The phase diagram of the ground state (in terms of the fractal dimension $\Gamma$) for different potential strength $\lambda$ and interaction $U$. 
The system size is $L=377$, and the NNN hopping strength is $\mu=0.35$. 
Here we set $t_1$ as the unit of energy. 
(b) The finite-size scaling of different regimes in the phase diagram. $m$ is the index of the Fibonacci number. 
In regime $\rm\uppercase\expandafter{\romannumeral1}$, we choose the point with parameters $U=-0.3,\lambda=2$ to perform the fitting. 
In regime $\rm\uppercase\expandafter{\romannumeral2}$, we choose $U=-0.3,\lambda=0.1$. In regime $\rm\uppercase\expandafter{\romannumeral3}$, we choose $U=0.3,\lambda=2$. 
In regime $\rm\uppercase\expandafter{\romannumeral4}$, we choose two points with $U=0.15,\lambda=0.1$ and $U=0.3,\lambda=0.421$. 
The solid lines show the fitting results.}
\end{figure}

We use the imaginary time evolution to numerically determine the mean-field ground state of the system $\ket{\psi_\text{GS}}$, which is determined by minimizing the energy $\expval*{H-\frac{U}{2}\sum_j|\psi_j|^2n_j}{\psi_{\text{GS}}}$. 
Figure~\ref{fig3}(a) shows the ground-state phase diagram of the system by plotting the fractal dimensions $\Gamma$ of $\ket{\psi_\text{GS}}$ for various potential strength $\lambda$ and interactions $U$ for a system with $L=F_{14} = 377$. 
Then we perform finite-size scaling to extrapolate the fractal dimensions in the thermodynamics limit~\cite{Yucheng_2022} by fitting the result to $\Gamma_m=\Gamma_{\infty}+b/m$, where $b$ and $F_{\infty}$ are the fitting parameters and $m$ is the Fibonacci index. 
For $\Gamma_{\infty}\rightarrow 0$ ($\Gamma_{\infty}\rightarrow 1$), the corresponding state is expected to be localized (extended), while for $0<\Gamma_{\infty}<1$, the state is critical. 
As shown in Fig.~\ref{fig3}(b), repulsive interactions ($U<0$) transform critical states into extended states (II) and drive localized states toward criticality (I), while attractive interactions ($U>0$) convert critical states into localized states (IV) with localized states remaining localized (III).

Interestingly, while both regimes III and IV are localized, they exhibit different behaviors. 
In particular, finite-size scaling indicates that the two regimes with different colors in regime $\rm\uppercase\expandafter{\romannumeral4}$ are localized states with different fitting coefficients $b$. 
Besides, it turns out that regime $\rm\uppercase\expandafter{\romannumeral3}$ retains inversion symmetry of the system while regime $\rm\uppercase\expandafter{\romannumeral4}$ breaks such a symmetry. 
This distinction can be fully understood by a perturbation theory analysis, which we relegate to the SM~\cite{SM}. 

\lettersection{Conclusion} 
In this work, we introduce a 1D quasiperiodic mosaic model with analytically solvable mobility edges that exhibit different phase transitions. 
By analyzing the Lyapunov exponent of the eigenstates, we rigorously demonstrate the existence of and derived exact analytical expressions for two distinct types of mobility edges: those separating extended and critical states, and those separating extended and localized states. 
Our numerical calculations of fractal dimensions confirm these analytical predictions. 
Furthermore, we propose a concrete experimental scheme based on Bose-Einstein condensates in optical lattices with engineered momentum-state transitions to realize our model. 
The effects of many-body interactions, studied under the mean-field approximation, reveal rich physics: repulsive interactions transform critical states into extended states and drive localized states toward criticality, while attractive interactions convert critical states into localized states with the emergence of two localized regimes distinguished by their symmetry properties.
Our findings provide a foundation for exploring the coexistence of different types of mobility edges in quasiperiodic systems and open new avenues for investigating the interplay between quasiperiodicity, localization, and many-body interactions in future theoretical and experimental studies.

\lettersection{Acknowledgments}
X.X. is supported by the National Science Foundation of China (Grant No.~12301218). 
X.D. is is supported by the Munich Quantum Valley. 
The work at Hong Kong is supported by the Research Grants Council of Hong Kong (Grants No.~CityU~11300421, CityU~11304823, and C7012-21G) and City University of Hong Kong (Project No. 9610428). K.H. is also supported by the Hong Kong PhD Fellowship Scheme. 

\bibliographystyle{apsrev4-2}
\bibliography{AME_ME_v4}

\end{document}


\global\long\def\id{\mathbbm{1}}
\global\long\def\ui{\mathbbm{i}}
\global\long\def\ud{\mathrm{d}}

\title{Supplemental Materials for ``One-dimensional quasiperiodic lattice with exactly different types of mobility edges''}

\author{Xu Xia}
\affiliation{Academy of Mathematics and Systems Science, Chinese Academy of Sciences, Beijing 100190, China }
\author{Weihao Huang}
\affiliation{Department of Physics, City University of Hong Kong, Kowloon, Hong Kong SAR}
\author{Ke Huang}
\affiliation{Department of Physics, City University of Hong Kong, Kowloon, Hong Kong SAR}
\author{Xiao Li}
\email{xiao.li@cityu.edu.hk}
\affiliation{Department of Physics, City University of Hong Kong, Kowloon, Hong Kong SAR}
\date{\today}

\maketitle

In this Supplemental Material, we first present some mathematical basis for computing the Lyapunov exponent in the main text. 
Then, we provide additional details of experimental realization and numerical simulation. 
Finally, we give details about the perturbative analysis of the symmetry-broken localized state shown in Fig.~3 of the main text.

\section{The computation of Lyapunov's exponent}
As discussed in the main text, we will only focus on the even sites of the lattice, with $j=2m$. 
The Lyapunov exponent (LE) $\gamma_\epsilon(E)$ can be calculated by taking the product of the transfer matrix $T_m(\theta)$, namely multiplying the transfer matrix $m$ times consecutively, which is written as 
\begin{align}
\mathcal{T}_m(\theta)=\prod_{k=0}^{m-1}T_k(\theta)
=\prod_{k=0}^{m-1}\frac{1}{t_2(2k-2)+t_1^2 / E}
\mqty[
E-2 t_1^2 / E-V_{2k} & -\left(t_2(2k)+t_1^2 / E\right) \\
\left(t_2(2k-2)+t_1^2 / E\right) & 0
],
\end{align}
where $V_{2k}$ is the quasiperiodic potential on site $2k$. 
The LE is then found as the limit of $\ln||\mathcal{T}_m(\theta)||/m$ as $m\to\infty$. 

The method that we use here to calculate the LE is the complexified phase approach. 
Specifically, by continuing the imaginary part of the phase $\epsilon$, we introduce a new LE,
\begin{align}
\gamma_\epsilon(E)=\lim_{m\rightarrow\infty}\frac{1}{m}\ln||\mathcal{T}_m(\theta+i\epsilon)||d\theta, \quad 
\mathcal{T}_m(\theta+ i\epsilon)=\prod_{k=0}^{m-1}T_k(\theta+ i\epsilon).
\end{align}

Relying on Avila's global theory, if we can obtain the LE $\gamma_\epsilon(E)$ when $\epsilon$ is sufficiently large, then we can trace back to the specific LE $\gamma_0(E)$ when $\epsilon=0$, namely the original LE.
To proceed, we first rewrite the transfer matrix as
\begin{equation}\label{eq3}
\begin{aligned}
T_0(\theta)&= \frac{1}{\left(2\mu \cos(2\pi (\theta-\omega/2))+t_1^2 / E\right)}
\mqty[
E-2 t_1^2 / E-2 \lambda \cos(2\pi \theta) & -\left(2\mu \cos(2\pi (\theta+\omega/2))+t_1^2 / E\right) \\
\left(2\mu \cos(2\pi (\theta-\omega/2))+t_1^2 / E\right) & 0
]\\
&=\frac{1}{\left(2\mu \cos(2\pi (\theta-\omega/2))+t_1^2 / E\right)}B_0(\theta),
\end{aligned}
 \end{equation}
where we have introduced the matrix $B_0(\theta)$ as follows:  
\begin{equation}
\begin{aligned}
 B_0(\theta)=
\mqty[
E-2 t_1^2 / E-2 \lambda \cos(2\pi \theta) & -\left(2\mu \cos(2\pi (\theta+\omega/2))+t_1^2 / E\right) \\
\left(2\mu \cos(2\pi (\theta-\omega/2))+t_1^2 / E\right) & 0
]. 
\end{aligned}
 \end{equation}
Then, $\gamma_\epsilon(E)$ can be expressed as
\begin{equation} \label{eq4}
\begin{aligned}
\gamma_{\epsilon}(E)=\lim_{m\rightarrow \infty}\frac{1}{m} \int\ln||\mathcal{B}_m(\theta +i\epsilon )|| d\theta - \int\ln \abs{\left(2\mu \cos(2\pi (\theta+i\epsilon))+t_1^2 / E\right)} \dd{\theta}
\equiv\gamma_{\epsilon}^B(E)+A_{\epsilon},
\end{aligned}
\end{equation}
where
\begin{equation}
\begin{aligned}
\gamma_{\epsilon}^B(E)=\lim _{m \rightarrow \infty} \frac{1}{m} \int \ln \left||\mathcal{B}_m(\theta+i \varepsilon)\right|| d \theta, \quad  
\mathcal{B}_m(\theta+i \epsilon)=\prod_{k=0}^{m-1} B_k(\theta+i \epsilon).
\end{aligned}
\end{equation}
From the above result, we can obtain 
\begin{equation}
A_{0}=\ln \qty[\frac{\sqrt{t_1^4-4 E^2 \mu^2}+t_1^2}{2|E|}]. 
\end{equation}
When $\epsilon \to +\infty$, a direct calculating result of $B_0(\theta+i\epsilon)$ gives
\begin{equation}\label{eq5}
B_0(\theta+i\epsilon)=e^{2\pi\epsilon}e^{i2\pi(\theta)}
\mqty[
 -\lambda  &  - \mu \\
\mu & 0 \\
] + \order{1}, 
\end{equation}
with $\|B_m(\theta+i\epsilon)\|=\|B_0(\theta+i\epsilon)\|$. 
Therefore, we find that 
\begin{align}
  \gamma^{B}_{\epsilon}(E)=2\pi\epsilon
  +\ln \qty[\frac{1}{2}\left(\lambda+\sqrt{\lambda^2-4 \mu^2}\right)]
  +\order{1}. 
\end{align}
As a function of $\epsilon$, $\gamma_{\epsilon}^B(E)$ is a convex, piecewise linear function whose slope is $2\pi$. 
Hence it is concluded that when $\epsilon$ goes to infinity, we obtain 
$${\gamma^{B}}_{\epsilon}(E)=2\pi\epsilon+\ln \qty[\frac{1}{2}\left(\lambda+\sqrt{\lambda^2-4 \mu^2}\right)].$$
Finally, according to Eq.~\eqref{eq4}, it leads to
$\gamma_0(E)=\max\{\alpha(E),0\}$, where 
\begin{align}
\alpha(E)=\ln \frac{|E|\left(\lambda+\sqrt{\lambda^2-4 \mu^2}\right)}{\sqrt{t_1^4-4 E^2 \mu^2}+t_1^2}.
\end{align}

\section{Experimental Realization}
\begin{figure}[htbp]
\begin{center}
\includegraphics[width=0.7\textwidth]{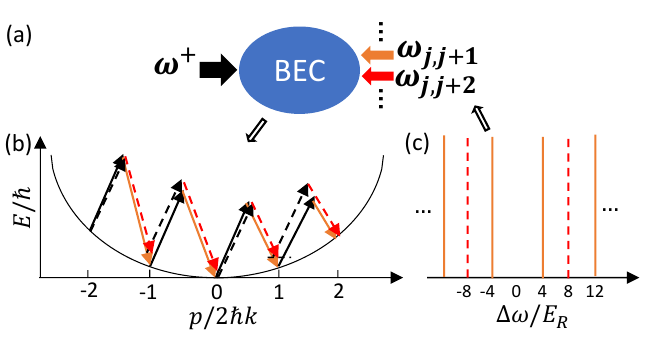}
\caption{(a) Illustration of the BEC platform to simulate the lattice model. (b) NN and NNN hoppings are realized by the two-photon (solid arrow) and four-photon (dashed arrow) Bragg transitions respectively. (c) The spectrum of the multi-frequency beam with frequency detuning $\Delta\omega$ is fed into the system to create 20 NN hoppings and 10 NNN hoppings.}
\label{fig_s3}
\end{center}
\end{figure}

In this section, we will elaborate on how to realize the lattice model using the BEC platform. 
The maximum achievable system size $L$ is 21 sites under current experimental conditions. 
The Bose-Einstein condensate (BEC) with $4\times 10^{4}$ $^{133}\text{Cs}$ atoms held in an optical trap, primarily formed by a single-frequency laser beam (wavelength $\lambda=\SI{1064}{nm}$, wavenumber $k=2\pi/\lambda$), is prepared shown in Fig.~\ref{fig_s3}(a). 
To construct a pair of counter-propagating laser beams, the primary beam is passed through two acousto-optic modulators (AOMs), where one AOM is driven with one single-frequency ($\omega_+$) radiofrequency (rf) tone and the other AOM is driven with the 20+10-frequency rf tones. 
The multifrequency beam is sent back to the same path as the primary single-frequency beam. 
The interference of the single-frequency and multifrequency beams can drive a series of Bragg transitions between two distanced momentum states, miming the hoppings between the lattice model as shown in Fig.~\ref{fig_s3}(b).

The recoil energy $E_R=\hbar^2k^2/2M_{\text{Cs}}=h\times \SI{5.3}{kHz}$, where $M_{\text{Cs}}$ is the mass of the atom Cs. 
The energy of the $21$ momentum states with momentum $p_j=2j\hbar k$ is $4E_Rj^2$ with $j=0,1,2,\cdots,20$.  

The $20$ NN hoppings between $p_j$ and $p_{j+1}$ can be realized by the two-photon Bragg transition. 
Specifically, the difference between the primary frequency and the corresponding frequency should satisfy
$$\Delta\omega_{j,j+1}=\omega_+-\omega_{j,j+1}=4E_R(2j+1)/\hbar-(\epsilon_{j+1}-\epsilon_{j})/\hbar.$$
Here $\epsilon_j=V_j$ is set up to realize the mosaic onsite quasiperiodic potential. Note that the detuning $\Delta$ of the laser from the atomic resonance is quite large. 
By tuning the strength ($\Omega_{j,j+1}$) and the phase ($\theta_{j,j+1}$) of each multifrequency component, one can tune the strength of NN hopping strength, which is estimated to be 
\begin{align}
t_1^{j,j+1}=\frac{\Omega_+\Omega_{j,j+1}}{\Delta}e^{i(\theta_+-\theta_{j,j+1})}.
\end{align}

The $10$ NNN hoppings are realized via a four-photon Bragg transition, where the difference between the primary frequency and the corresponding frequency should satisfy 
$$\Delta\omega_{j,j+2}=\omega_+-\omega_{j,j+2}=4E_R(2j+2)/\hbar-(\epsilon_{j+2}-\epsilon_j),$$
with even $j$. 
Similarly, by tuning the strength ($\Omega_{j,j+2}$) and the phase ($\theta_{j,j+2}$) of each multifrequency component, one can tune the strength of NNN hopping strength, which is estimated to be 
\begin{align}
t_2^{j,j+1}=\frac{\Omega_+^2\Omega^2_{j,j+2}}{\Delta^2(4E_R)}e^{i(2\theta_+-2\theta_{j,j+2})}.
\end{align}
The frequency spectrum simulating hoppings between lattice sites is shown in Fig.~\ref{fig_s3}(c).

\section{Numerical Simulation}
\begin{figure}[b]
\centering
\includegraphics[width=0.5\textwidth]{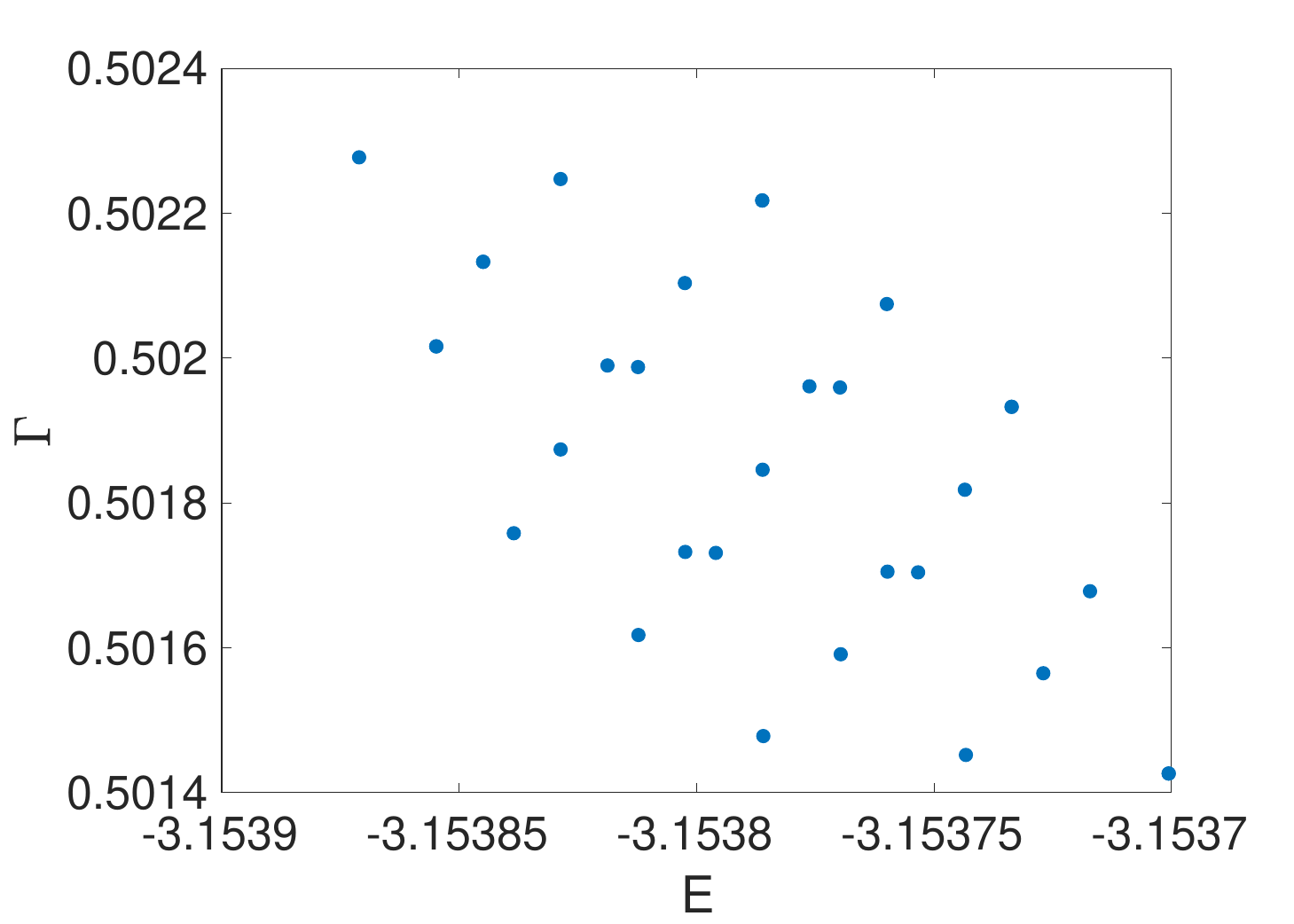}
\caption{The fractal dimension versus the energy of 5000 different random initial states. $L=377$, $\mu=0.35$, $\lambda=1.2632$, $U=-1$.}\label{figs1} 
\end{figure}

In this section, we discuss the numerical simulation of the interacting ground state. We perform imaginary time evolution to find the ground state of the interacting system. 
For one Hamiltonian, we choose different initial states to avoid local minima. 
For each realization of the initial state, the energy uncertainty $\sqrt{\langle H^2\rangle-\langle H\rangle^2}$ is on the order of $10^{-15}$. 
For system size $L=377$, when $\lambda=1.2632$, $U=-1$, we perform $5000$ random intial states. 
As shown in Fig.~\ref{figs1}, the maximum difference $\Gamma_{max}-\Gamma_{min}\approx 8.5\times 10^{-4}$, for which the influence of local minimum on the fractal dimension is neglected. 
Therefore, we use several random initial states to construct the phase diagram for $L=377$ and do the finite-size scaling. 
However, we must find the global minima with enough random initial states to distinguish the inversion symmetry in different regimes. 

\begin{figure}[t]
\centering
\includegraphics[width=0.8\textwidth]{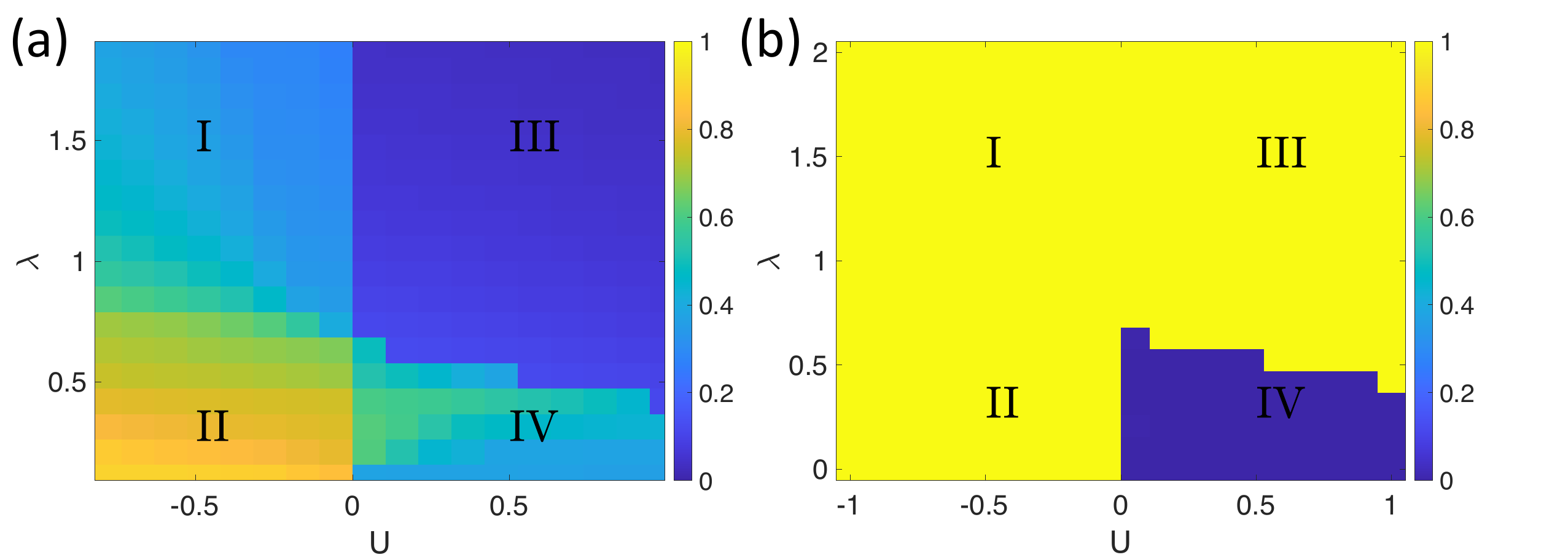}
\caption{
 (a) Fractal dimension $\Gamma$ of different eigenstates as a function of the interaction strength $U$ and quasiperiodic potential strength $\lambda$ with size $L=89$. Here, the hopping strength $t_1$ has been set as the unit
of energy and $\mu=0.35$. (b) Phase diagram for the expectational values of the inversion symmetry of GS.}
\label{fig_s2}
\end{figure}

Fig.~\ref{fig_s2}(a) shows the phase diagram with the system size $L=89$, which costs less time and different initial states to find the GS of the system. Fig.~\ref{fig_s2}(b) shows the corresponding phase diagram about the expectational value of the inversion symmetry $\hat{P}$ of the GS, i.e., $\bra{GS}\hat{P}\ket{GS}$. 

\section{Perturbation theory analysis of the phase diagram}
In this section, we provide a detailed calculation of the perturbation theory. Supposing that the localized state localizes at site $j$ and treating the hopping and the on-site potential as perturbations, the eigenvalue of the eigenstate $\ket{\psi^{(0)}}=\ket{j}$ of the unperturbed Hamiltonian, $-U\sum_n|\psi_n^{(0)}|^2n_n$, is $E^{(0)}_j=-U$. At the same time, we have $E^{(0)}_n=0$ for $n\neq j$. By performing perturbation theory, the eigenstate of the mean-field Hamiltonian is approximated by
\begin{align}
\ket{\psi_{\text{approx}}}=&\ket{\psi^0}+\ket{\psi^1}+\ket{\psi^2_j}\nonumber\\
\ket{\psi^0}=&\ket{j}\nonumber\\
\ket{\psi^1}=&\sum_{n\neq j}\frac{\bra{n}H\ket{j}}{E^{(0)}_j-E^{(0)}_n}\ket{n} 
=-\frac{t_1}{U}\ket{j-1}-\frac{t_1}{U}\ket{j+1}-\frac{t_2(j-2)}{U}\ket{j-2}-\frac{t_2(j)}{U}\ket{j+2}\nonumber\\
\ket{\psi^2_j}=&-\frac{1}{2}\ket{j}\sum_{n\neq j}\frac{|\bra{n}H\ket{j}|^2}{(E_j^{(0)}-E^{(0)}_n)^2}
=-\left[\frac{t_1^2}{U^2}+\frac{t_2^2(j-2)+t_2^2(j)}{2U^2}\right]\ket{j},
\end{align}
where $\ket{\psi^i}$ is the $i\text{th}$ order perturbation wavefunction with respect to interaction strength $U$. 
$\ket{\psi^2_j}$ is the second-order perturbation wavefunction component located on site $j$. We will show afterward that the other components of the second-order correction to the wavefunction $\ket{\psi^2_{n\neq j}}$ do not contribute to the energy up to the second order $\order{U^{-2}}$. The approximated energy of the system up to the order of $\order{U^{-2}}$ is
\begin{align}
E_{\text{approx}}=&\bra{\psi_{\text{approx}}} H-\frac{U}{2}\sum_n|\psi_n^{\text{approx}}|^2n_n\ket{\psi_{\text{approx}}}\nonumber\\
=&-\frac{U}{2}-\frac{2t_1^2}{U}-\frac{t_2^2(j)+t_2^2(j-2)}{U}+\frac{2[t_1^2t_2(j)+t_1^2t_2(j-2)]}{U^2}+V_j\left[1-\frac{2t_1^2}{U^2}-\frac{t^2_2(j-2)}{U^2}-\frac{t^2_2(j)}{U^2}\right]\nonumber\\
&+\frac{t_1^2(V_{j-1}+V_{j+1})}{U^2}+\frac{t_2^2(j-2)V_{j-2}+t^2_2(j)V_{j+2}}{U^2}+\order{U^{-2}}.\label{E_loc}
\end{align}
In the following, we prove why the precision of the approximated energy is in the order of $\order{U^{-2}}$. 
The error of the energy, $\xi_{E}$ is defined as 
\begin{align}
\xi_E=&\bra{\psi_{\text{exact}}} H-\frac{U}{2}\sum_n|\psi^{\text{exact}}_n|^2n_n\ket{\psi_{\text{exact}}}-\bra{\psi_{\text{approx}}} H-\frac{U}{2}\sum_n|\psi^{\text{approx}}_n|^2n_n\ket{\psi_{\text{approx}}}\nonumber\\
=&-\bra{\psi_{\text{approx}}} \frac{U}{2}\sum_n(|\psi^{\text{exact}}_n|^2-|\psi^{\text{approx}}_n|^2)n_n\ket{\psi_{\text{approx}}}+\bra{\delta\psi} H-\frac{U}{2}\sum_n|\psi^{\text{exact}}_n|^2n_n\ket{\psi_{\text{approx}}}\nonumber\\
&+\bra{\psi_{\text{approx}}} H-\frac{U}{2}\sum_n|\psi^{\text{exact}}_n|^2n_n\ket{\delta\psi}+\bra{\delta\psi} H-\frac{U}{2}\sum_n|\psi^{\text{exact}}_n|^2n_n\ket{\delta\psi},\label{error}
\end{align}
where the exact wavefunction is $\ket{\psi_{\text{exact}}}=\ket{\psi_{\text{approx}}}+\ket{\delta\psi}$, and we have 
$$|\psi^{\text{exact}}_n|^2=|\psi^{\text{approx}}_n+\delta\psi_n|^2=|\psi^{\text{approx}}_n|^2+|\delta\psi_n|^2+2\delta\psi_n\psi^{\text{approx}}_n.$$ 
The error of the approximated wavefunction $\ket{\delta\psi}$ is estimated as follows: 
\begin{align}
\ket{\delta\psi}=&\ket{\psi^2_{n\neq j}}+\ket{\psi^3}+\order{U^{-4}},\nonumber\\
\ket{\psi^2_{n\neq j}}=&\sum_{m\neq j}\sum_{n\neq j}\ket{m}\frac{\bra{m}H\ket{j}\bra{n}H\ket{j}}{(E^{(0)}_j-E^{(0)}_m)(E^{(0)}_j-E^{(0)}_n)}-\sum_{n\neq j}\ket{n}\frac{\bra{n}H\ket{j}\bra{j}H\ket{j}}{(E^{(0)}_j-E^{(0)}_n)^2}\nonumber\\
=&\ \frac{1}{U^2}(t^2_1\ket{j-2}+t^2_1\ket{j+2}+t_1t_2(j-2)\ket{j-1}+t_1t_2(j)\ket{j+1})\nonumber\\
&-\frac{V_j}{U^2}(t_1\ket{j-1}+t_1\ket{j+1}+t_2(j-2)\ket{j-2}+t_2(j)\ket{j+2}),\nonumber\\
\ket{\psi^3_j}=&\sum_{k_1\neq j}\sum_{k_2\neq j}\ket{j}\left[-\frac{\bra{j}H\ket{k_2}\bra{k_2}H\ket{k_1}\bra{k_1}H\ket{j}+\bra{k_2}H\ket{j}\bra{k_1}H\ket{k_2}\bra{j}H\ket{k_1}}{2(E^{(0)}_j-E^{(0)}_{k_2})^2(E^{(0)}_j-E^{(0)}_{k_1})}+\frac{|\bra{j}H\ket{k_1}|^2\bra{j}H\ket{j}}{(E^{(0)}_j-E^{(0)}_{k_1})^3}\right] \nonumber \\
=&\frac{1}{U^3}[2t_1^2t_2(j-2)+2t_1^2t_2(j)-2t_1^2V_j-t_2^2(j-2)V_j-t_2^2(j)V_{j}]\ket{j}\nonumber\\
=&\psi^3_j\ket{j}. 
\end{align}
In the following, we will show that $\ket*{\psi^3_{n\neq j}}$ gives no contribution to the energy up to the order of $\order{U^{-2}}$.

Now let us estimate each term in Eq.~\eqref{error} up to the order of $\order{U^{-2}}$ one by one. 
The first term in Eq.~\eqref{error} is
\begin{align}
&-\bra{\psi_{\text{approx}}} \frac{U}{2}\sum_n(|\psi^{\text{exact}}_n|^2-|\psi^{\text{approx}}_n|^2)n_n\ket{\psi_{\text{approx}}} \\
=& -\bra{\psi^{\text{approx}}} \frac{U}{2}\sum_n|\delta\psi_n|^2n_n\ket{\psi^{\text{approx}}}
-\bra{\psi^{\text{approx}}} \frac{U}{2}\sum_n2\delta\psi_n\psi_n^{\text{approx}}n_n\ket{\psi^{\text{approx}}}\nonumber \\
=&\;\order{U^{-2}}-\bra{\psi^{\text{approx}}} U\sum_n\delta\psi_n\psi_n^{\text{approx}}n_n\ket{\psi^{\text{approx}}},
\end{align}
where
\begin{align}
-\bra{\psi^{\text{approx}}} U\sum_n\delta\psi_n\psi_n^{\text{approx}}n_n\ket{\psi^{\text{approx}}}=&-\bra{\psi^0}U\sum_{n\neq j}\psi^2_{n\neq j} \psi_n^{\text{approx}}n_n \ket{\psi^0}-\bra{\psi^0}U\sum_n\psi^3_nn_n \psi_n^0 \ket{\psi^0}+\order{U^{-2}}\nonumber \\
=&-\bra{j}U\sum_{n\neq j}\psi^2_{n\neq j} \psi_n^{\text{approx}}n_n \ket{j}-\bra{j}U\sum_n\psi^3_n \psi_n^0 n_n\ket{j}+\order{U^{-2}} \nonumber \\
=&\ 0-\bra{j}U\psi^3_j\psi^0_j n_j\ket{j}-\bra{j}U\sum_{n\neq j}\psi^3_n \psi_n^0 n_n\ket{j}+\order{U^{-2}}\nonumber\\
=&\ 0-\bra{j}U\psi^3_j\psi^0_j \ket{j}-0+\order{U^{-2}}\nonumber\\
=&-\frac{2}{U^2}(t_1^2t_2(j-2)+t_1^2t_2(j)-V_jt_1^2-V_jt_2^2(j-2)/2-V_jt_2^2(j)/2)+\order{U^{-2}}\nonumber\\
=&-U\psi_j^3+\order{U^{-2}}.\label{term1}
\end{align}
The second term in Eq. \eqref{error} is
\begin{align}
\bra{\delta\psi}& H-\frac{U}{2}\sum_n|\psi^{\text{exact}}_n|^2n_n\ket{\psi_{\text{approx}}} \notag\\
=&\bra{\delta\psi}H\ket{\psi^0}-\bra{\delta\psi}\frac{U}{2}\sum_n|\psi^{\text{exact}}_n|^2n_n\ket{\psi^0}-\bra{\delta\psi}\frac{U}{2}\sum_n|\psi_n^{\text{exact}}|^2n_n\ket{\psi^1}+\order{U^{-2}},\label{s12}
\end{align}
where
\begin{align}
\bra{\delta\psi}H\ket{\psi^0}=&\bra{\psi_{n\neq j}^2}H\ket{\psi^0}+\order{U^{-2}}\nonumber\\
=&\;\frac{2}{U^2}[t_1^2t_2(j)+t_1^2t_2(j-2)-V_jt_1^2-V_jt_2^2(j-2)/2-V_jt_2^2(j)/2]+\order{U^{-2}}\nonumber\\
=&\ U\psi_j^3+\order{U^{-2}}.\label{s13}
\end{align}
Because 
\begin{align*}
  \sum_n|\psi^{\text{exact}}_n|^2n_n=\sum_n(|\psi^{0}_n|^2+|\psi^{1}_n|^2+\psi^{0}_n\psi^1_n+2\psi^0_n\psi^2_n+\order{U^{-2}})n_n=|\psi^{0}_j|^2n_j+2\psi^0_j\psi^2_jn_j+\sum_{n\neq j}(|\psi^{1}_n|^2+\order{U^{-2}})n_n
\end{align*}
and $\bra{\delta \psi}\ket{\psi^0}=\left[\bra*{\psi^2_{n\neq j}}+\bra*{\psi^3_{n\neq j}}+\bra{\psi^3_{j}}\right]\ket{j}=\bra{\psi^3_{j}}\ket{j}$, the second term in Eq. \eqref{s12} is
\begin{align}
-\bra{\delta\psi}\frac{U}{2}\sum_n|\psi^{\text{exact}}_n|^2n_n\ket{\psi^0}=&-\bra{\delta\psi}\frac{U}{2}(|\psi^{0}_j|^2+2\psi^0_j\psi^2_j)n_j\ket{j} -\bra{\delta\psi}\frac{U}{2}\sum_{n\neq j}|\psi^{1}_n|^2n_n\ket{j}+\order{U^{-2}}\nonumber \\
=&-\frac{U}{2}(1+2\psi^2_j)\bra{\psi^3_j}\ket{j}+\order{U^{-2}}, 
\label{s14}
\end{align}
where the second term in the second to last step vanishes. 
Because $\frac{U}{2}2\psi^2_j\bra{\psi^3_j}\ket{j}=\order{U^{-2}}$, the above equation becomes
\begin{align}
-\frac{U}{2}\bra{\psi^3_j}\ket{j}+\order{U^{-2}}
=-\frac{U}{2}\psi_j^3+\order{U^{-2}}.
\end{align}
The third term in Eq. \eqref{s12} is 
\begin{align}
-\bra{\delta\psi}\frac{U}{2}\sum_n|\psi_n^{\text{exact}}|^2n_n\ket{\psi^1}=&-\bra{\delta\psi}\frac{U}{2}\sum_n|\psi_n^{0}|^2n_n\ket{\psi^1_{n\neq j}}+\order{U^{-2}}\nonumber\\
=&-\bra{\delta\psi}\frac{U}{2}|\psi_j^{0}|^2n_j\ket{\psi^1_{n\neq j}}+\order{U^{-2}}
=\order{U^{-2}},\label{s17}
\end{align}
where the first term in the second to last step vanishes. 
We then substitute Eq.~\eqref{s13}, \eqref{s14}, \eqref{s17} into Eq.~\eqref{s12}, and find that 
\begin{align}
\bra{\delta\psi} H-\frac{U}{2}\sum_n|\psi^{\text{exact}}_n|^2n_n\ket{\psi_{\text{approx}}}=\frac{U}{2}\psi_j^3+\order{U^{-2}}.\label{term2}
\end{align}
Similarly, we get the third term in Eq. \eqref{error}
\begin{align}
\bra{\psi_{\text{approx}}} H-\frac{U}{2}\sum_n|\psi^{\text{exact}}_n|^2n_n\ket{\delta\psi}=\bra{\delta\psi} H-\frac{U}{2}\sum_n|\psi^{\text{exact}}_n|^2n_n\ket{\psi_{\text{approx}}}=\frac{U}{2}\psi_j^3+\order{U^{-2}}.\label{term3}
\end{align}
The fourth term of Eq.~\eqref{error} is
\begin{align}
\bra{\delta\psi} H-\frac{U}{2}\sum_n|\psi^{\text{exact}}_n|^2n_n\ket{\delta\psi}
=&\order{U^{-4}}-\bra{\delta\psi}\frac{U}{2}\sum_n|\psi^{\text{exact}}_n|^2n_n\ket{\delta\psi}\nonumber\\
=&\order{U^{-4}}-\bra{\psi^2_{n\neq j}}\frac{U}{2}\sum_n|\psi^{0}_n|^2n_n\ket{\psi^2_{n\neq j}}\nonumber\\
=&\order{U^{-4}}-\bra{\psi^2_{n\neq j}}\frac{U}{2}|\psi^{0}_j|^2n_j\ket{\psi^2_{n\neq j}}
=\order{U^{-4}},\label{term4}
\end{align}
where the second term in the second to last step vanishes. 
Combining Eq.~\eqref{term1}, \eqref{term2}, \eqref{term3}, and \eqref{term4}, we find that the error of the energy is $\xi_E = \order{U^{-2}}$.

After minimizing Eq.~\eqref{E_loc}, we know where the localized state is located.


